# Sub-micron diameter micropillar cavities with high Quality factors and ultra-small mode volumes

Yinan Zhang,[*] Marko Lončar

*School of Engineering and Applied Sciences, Harvard University, 33 Oxford Street, Cambridge, MA 02138, USA*

[*]*Corresponding author: yinan@seas.harvard.edu*

We theoretically demonstrate high Quality factor ($Q \sim 3 \times 10^6$) micropillar cavities with record low mode volumes [$V \sim 0.1(\lambda/n)^3$] based on the $TiO_2/SiO_2$ material system. The proposed cavities have $Q/V$ three orders of magnitude larger than previously reported ones. We show that our cavity embedded with diamond nanocrystal provides a feasible platform for cavity quantum electrodynamics experiments in the strong coupling limit. 2008 Optical Society of America

*OCIS codes:* 140.3948, 270.0270, 140.7260.

Micropillar optical cavities, typically used in low-threshold vertical-cavity surface-emitting lasers (VCSELs) [1-3], have recently attracted considerable attention as a promising platform for solid-state implementations of cavity quantum electrodynamics (cQED) experiments [4, 5]. These applications benefit from large quality factor ($Q$) that can be obtained in micropillar cavities, which in turn results in a long photon life time $\kappa = \omega/2Q$ ($\omega$ is the radial frequency of the cavity mode). For example, high-$Q$ micropillar cavities ($Q \sim 165,000$) have been demonstrated recently for large diameter ($d=4\mu m$) pillars, resulting in a relatively large mode volume [$V>50(\lambda/n)^3$] [6]. However, for applications in cQED, coupling between an emitter and a photon localized in the cavity requires a large Rabi frequency $g$ that is proportional to $1/\sqrt{V}$. In order to decrease the mode volume, it is necessary to decrease the pillar diameter and thereby improve the radial confinement of light. However, this can lead to a significant reduction in the $Q$ factor in the traditional micropillar designs [7], and the best $Q$ factors reported in the case of sub-micron diameter micropillars have been theoretically limited to approximately 2,000 [8].

In this letter, we for the first time theoretically demonstrate sub-micron diameter micropillar cavities with an ultra-high $Q/V$ that is three orders of magnitude larger than previously reported. This is achieved by simultaneously increasing the $Q$ ($Q\sim 3\times 10^6$) and reducing the mode volume [$V\sim 0.1(\lambda/n)^3$], using a bandgap tapering method developed recently in 1D photonic crystal structures [9, 10]. While the proposed approach is general and can be applied to a range of material systems and applications, our cavities are designed to operate at 637nm wavelength and therefore are suitable for coupling to nitrogen-vacancy (NV) color centers embedded within diamond nanocrystals positioned at the middle of the cavity. NV color centers have recently attracted significant attention as promising quantum emitters [11]. NVs emission is broad band (630nm-750nm) with stable zero-phonon line at 637nm visible even at room temperature.

Our micropillar is based on two distributed Bragg mirrors (DBR) that consists of $TiO_2$ ($n_{TiO2}=2.4$) and $SiO_2$ ($n_{SiO2}=1.5$) alternating layers, and a $TiO_2$ spacer of thickness $s$ sandwiched between them [Fig. 1(a)]. The micropillar cavity $Q$ is inversely proportional to the cavity losses that in turn can be separated into two components: the transmission losses due to the finite length of the DBRs, and the scattering losses at the spacer/mirror interfaces. The former can be minimized (for a given number of $TiO_2/SiO_2$ pairs) using a "quarter stack" DBR that consists of $TiO_2/SiO_2$ layers with thickness $\lambda/4n_{eff}$, where $\lambda=637$nm. This also maximizes light confinement along the pillar axis, resulting in a minimized mode volume. For example, in the case of a micropillar with diameter $d=340$nm, the thicknesses of $TiO_2$ and $SiO_2$ layers in DBR are 82.6nm and 147.4nm, respectively, resulting in total DBR periodicity of $a=230$nm. Next, the thickness of the spacer is chosen ($s=220$nm) in order to position the cavity resonance at 637nm, and the quality factor ($Q$) of such cavity is obtained using finite-difference time-domain method (FDTD). We find $Q\sim100$ which is consistent with previous reports [7]. The low $Q$ is attributed to scattering losses arising from the mode profile mismatch between the localized cavity mode and evanescent Bloch mode inside the DBRs [12]. Fig. 1(a) illustrates the profile mismatch for $E_r$ component. To suppress this mode mismatch and the resulting scattering losses, we use the mode matching technique previously developed for 1D photonic crystal cavities [9]. We substitute the uniform center segment with a single $TiO_2/SiO_2$ pair with the same aspect ratio as in the DBR but a smaller thickness $w$. When $w=0.67a$, the cavity resonates at $\lambda=637$nm with $Q=6,000$, a 60-fold improvement over the conventional design.

In order to increase the $Q$ factor further, we incorporate more $TiO_2/SiO_2$ segments with varying the thicknesses $w_i$ ($i$ is the segment number). This can also be seen as a "tapered DBR" approach, where each taper section further reduces the mode mismatch in order to suppress the scattering losses. In Fig. 2(a), we use 4 tapered segments and 20 DBR pairs at each side. In order to set the resonating wavelength at 637nm, the thickness of each taper segment is precisely tuned to $w_1=215.3$nm, $w_2=202.4$nm, $w_3=191.0$nm, and $w_4=180.8$nm. The resulting mode has a $Q$ factor of 250,000 and a mode volume of $0.07(\lambda/n)^3$, which represents at least three orders of magnitude enhancement of $Q/V$ compared to any previous micropillar designs. As shown in Fig. 2(b), this

high-$Q$ mode has an anti-node at the central $TiO_2$ segment and therefore is ideally suited for coupling to quantum emitters, such as an NV center in diamond or a semiconductor nanocrystal, embedded within this layer. We also find a second-order cavity mode [Fig. 2(c)] at wavelength of 685nm with a respectable $Q$=110,000. The $Q$ factor of the fundamental mode can be enhanced by increasing the tapering process. For instance, we obtain $Q$=3,000,000 and $V$=0.10$(\lambda/n)^3$ with 10 taper segments. However, increasing the cavity length pulls higher-order modes from the dielectric band into the bandgap, as shown in Fig. 2(d) and (e). These higher order modes can potentially couple to the emitter placed at the center of the cavity and the resulting multi-mode cavity is not desirable. Therefore from here on, we only consider 4-taper-segment cavities, which limits our $Q$ to 250,000.

Next, we optimize the diameter of our micropillar cavity to minimize its mode volume. It can be seen in Fig. 3(a) that the smallest mode volume is obtained at $d$=340nm. For $d$<340nm, $V$ increases due to the reduced confinement in the axial direction: the effective mode index contrast between $TiO_2$ and $SiO_2$ is reduced for small diameters, and therefore the width of the bandgap decreases, resulting in deeper penetration of the cavity mode into the DBRs, thus increasing $V$. For $d$>340nm, however, the mode is almost completely confined within the pillar and the $n_{eff}$ of each segment approach the refractive index of the material. Therefore, the width of the bandgap remains approximately constant as the pillar diameter increases and the axial confinement remains the same. However, $V$ increases due to the larger mode cross-section (radial confinement increases). The trade-off between radial and axial confinement results in an optimized diameter of $d$=340nm. For the 4-taper-segment cavity, we also find the cavity modes with $TE_{01}$ and $TM_{01}$ profiles at wavelengths of 578nm and 492nm, respectively. Both of these modes peak at the central $TiO_2$ segment. However, the lateral electric field density profiles shown in Fig. 3(b) indicate that these modes have a node at the center of the micropillar and therefore will not couple efficiently to the nano-emitter placed at the center. Moreover, these modes are detuned from the emission spectrum of an NV center and therefore are not of interest. It is also important to note that if the pillar diameter increases further, additional higher-order modes, with the same azimuthal order as $HE_{11}$ (e.g. $EH_{11}$), are allowed. These modes can couple to the fundamental $HE_{11}$ cavity mode and thus introduce additional loss mechanism [7], and reduce the $Q$ factor of the fundamental cavity mode.

Finally, we investigate the potential of our microcavity for applications in cQED. Diamond nanocrystals (size<50nm) with single NV color centers could be embedded within the central $TiO_2$ layer in the cavity. It is interesting to note that refractive index of $TiO_2$ is very close to that of diamond, and therefore the amount of light scattering at the interface between diamond and $TiO_2$ is negligible. The light-matter interaction process can be characterized by the coupling strength (Rabi frequency) $g$, the emitter spontaneous emission rate $\gamma$, and the photon loss rate $\kappa$. The common emission life time of a diamond NV center is $\tau \approx$ 20ns, leading to $\gamma = 2\pi/\tau = 2\pi \times 0.05 GHz$. The photon loss rate $\kappa$ is evaluated via $\kappa = \omega/2Q = 2\pi \times 0.94 GHz$ ($Q$=250,000). Assuming the emitter is positioned at the mode maximum, and its dipole moment is parallel to the field vector, coupling strength can be obtained as $g_0 = \sqrt{\dfrac{3\pi c^3}{2\tau\omega^2 n_e n_c^2 V}}$, where $V$ is the mode volume of the cavity, and $n_e$ and $n_c$ are the refractive indices of the emitter and cavity, respectively. In a diamond NV center, the zero-phonon line only contributes about 5% of the total emission, the rest being emitted into the phonon sideband. Therefore, only $\approx$ 5% of the total emission is coupled to the cavity mode, so $g$ must be scaled by a factor of $1/\sqrt{20}$, resulting in $g = g_0/\sqrt{20} = 2\pi \times 7.1 GHz$. By comparing the characteristic frequencies, we conclude that in our system, $g > \kappa, \gamma$, and that the system is well into the strong coupling regime. Furthermore, it is interesting to note that the system is in the strong coupling regime even when $Q>30,000$, which can be obtained with only 10 pairs of DBR mirrors on each side of the tapered section.

In conclusion, we have demonstrated that high Quality factor micropillar cavities can be realized with sub-micron diameter pillars. We have engineered the cavities with a record low mode volume of $V=0.07(\lambda/n)^3$ and a Quality factor of 250,000. We expect, however, that realistic fabricated structures will have reduced quality factors due to fabrication-related imperfections, including surface roughness, slanted walls and material absorption. One possible approach to overcome these problems is based on oxide-aperture design [13]. We predicted that by embedding a diamond nanocrystal with an NV color center at the middle of the cavity, the strong coupling limit of light-matter interaction can be achieved. Our method can be easily adapted to different material systems and enable realization of an ultra-high $Q/V$ cavity in AlAs/GaAs platform suitable for realization of low-threshold VCSELs, for example.

This work was supported in part by NSF grant ECCS – 0708905 "NIRT: Photon and Plasmon Engineering in Active Optical Devices based on Synthesized Nanostructures" and Harvard NSEC center. The authors acknowledge the assistance provided by Murray W. McCutcheon. Yinan Zhang would also like to dedicate this work to Honglie Yin and Erfang Zhang.

**Figures**

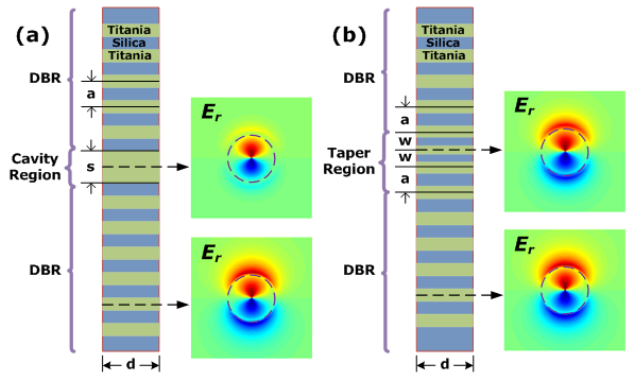

Fig. 1. (a) Traditional design of micropillar cavities and (b) modified design where the center segment is substituted by titania/silica pairs. The lateral mode profile of $E_r$ component for cavity mode and evanescent Bloch mode that exists inside DBRs are shown on the right of the cavity layout. Improved mode-matching can be seen in (b).

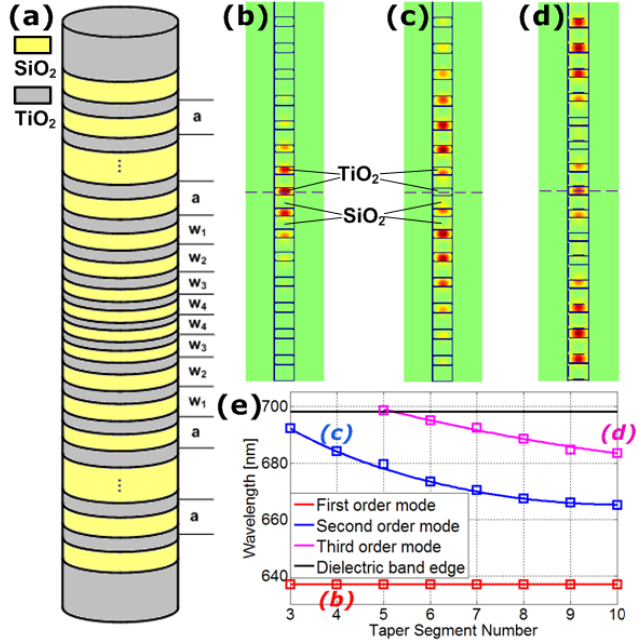

Fig. 2. (a) Schematic of a 4-taper-segment micropillar cavity. (b)(c) Electric field density profile of the first and second order mode, respectively. (d) Electric field density profile of the third

order mode of the 10-taper-segment micropillar cavity. (e) Mode diagram as a function of taper segment number.

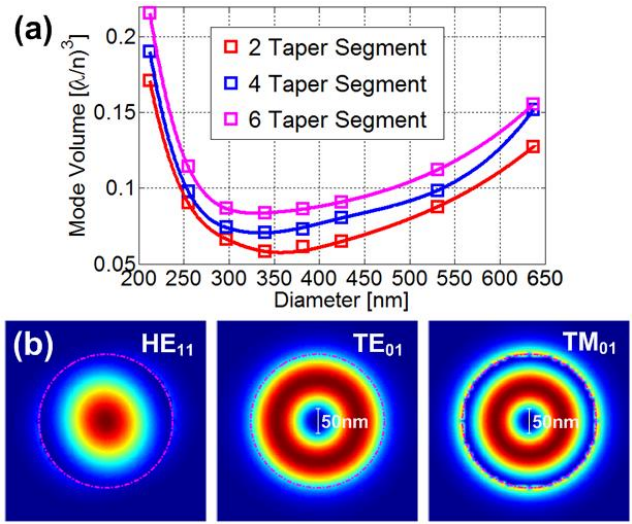

Fig. 3. (a) Mode volume as a function of micropillar diameter. Here all the modes are first-order $HE_{11}$ modes resonating at 637nm. (b) Lateral electric field density profiles of $HE_{11}$ ($\lambda$=637nm), $TE_{01}$ ($\lambda$=578nm) and $TM_{01}$ ($\lambda$=492nm) cavity modes.

**List of figure captions**

1. (a) Traditional design of micropillar cavities and (b) modified design where the center segment is substituted by titania/silica pairs. The lateral mode profile of $E_r$ component for cavity mode and evanescent Bloch mode that exists inside DBRs are shown on the right of the cavity layout. Improved mode-matching can be seen in (b).

2. (a) Schematic of a 4-taper-segment micropillar cavity. (b)(c) Electric field density profile of the first and second order mode, respectively. (d) Electric field density profile of the third order mode of the 10-taper-segment micropillar cavity. (e) Mode diagram as a function of taper segment number.

3. (a) Mode volume as a function of micropillar diameter. Here all the modes are first-order $HE_{11}$ modes resonating at 637nm. (b) Lateral electric field density profiles of $HE_{11}$ ($\lambda$=637nm), $TE_{01}$ ($\lambda$=578nm) and $TM_{01}$ ($\lambda$=492nm) cavity modes.